\documentclass[12pt,a4paper]{article}

\pdfoutput=1
\usepackage[dvips]{graphicx} 
\usepackage{bmpsize}
\usepackage{datetime}
\usepackage{amsmath,amssymb,array,calc,rotating,epsfig,psfrag, amscd}

\usepackage{color}
\usepackage{verbatim}
\usepackage[colorlinks=false,
    linktoc=all, 
    pdfstartview=FitV,
    bookmarksopen=true]{hyperref}
\usepackage[left=2cm,top=1cm,right=3cm,nohead]{geometry}
\usepackage[english]{babel}

\usepackage{cite}

\renewcommand{\d}{\textrm{d}}

\newcommand{\bea}{\begin{eqnarray}}
\newcommand{\eea}{\end{eqnarray}}
\newcommand{\be}{\begin{equation}}
\newcommand{\ee}{\end{equation}}

\definecolor{cardinal}{rgb}{0.6,0,0}
\definecolor{darkgreen}{rgb}{0,0.5,0}
\definecolor{golden}{rgb}{0.92, 0.7, 0}
\definecolor{midnight}{rgb}{0, 0, 0.5}
\definecolor{darkblue}{rgb}{0.2, 0, 0.8}

\newcommand{\beq}{\begin{equation}\begin{aligned}}
\newcommand{\eeq}{\end{aligned}\end{equation}}

\numberwithin{equation}{section}

\begin{document}
\begin{flushright}
\small
UUITP-09/18
\normalsize
\end{flushright}

\thispagestyle{empty}
\vspace{1cm}
\begin{center}
\baselineskip=13pt {\LARGE \bf{ What if string theory has no de Sitter vacua?}}
 \vskip1.5cm 
{\large Ulf H. Danielsson$^a$
 and Thomas Van Riet$^b$}  \\
\vskip0.5cm
 \textit{$^b$ Institutionen f{\"o}r fysik och astronomi,\\ Uppsala Universitet, Uppsala, Sweden}\\
\href{mailto:ulf.danielsson@physics.uu.se}{ulf.danielsson @ physics.uu.se}

\vskip0.5cm
 \textit{$^c$Instituut voor Theoretische Fysica, K.U. Leuven,\\
  Celestijnenlaan 200D, B-3001 Leuven, Belgium   }\\
\href{mailto:thomas.vanriet @ fys.kuleuven.be}{thomas.vanriet @ fys.kuleuven.be}
\vskip3.5cm
\end{center}

\begin{abstract}
\noindent
We present a brief overview of attempts to construct de Sitter vacua in string theory and explain how the results of this 20-year endeavor could point to the fact that string theory harbours no de Sitter vacua at all. Making such a statement is often considered controversial and ``bad news for string theory". We discuss how perhaps the opposite can be true. 
\end{abstract}

\clearpage
\tableofcontents

\newpage

\section{Introduction}\label{sec:intro}

Since the middle of the 1970s string theory has been argued to be the way in which quantum mechanics and general relativity are to be unified into a theory of quantum gravity. Encouraged by the progress and challenges in particle physics at the time, there were wide spread hopes that string theory could be developed into a theory of everything, capable of explaining and deriving all fundamental properties of the laws of physics. Through string theory, the nature of the fundamental forces and particle spectra, together with coupling constants and particle masses, would be uniquely determined by mathematical and logical consistency. That was at least the dream and promise that was presented to the general public. 

The focus of the period was on high energy and small scales, including the nature of the extra dimensions and their importance for physics accessible to experiments. Most string theorists had a background in high energy physics, and cosmology did not play an important role for the subject. Precision cosmology was in its infancy, and inflation was not the talk of the town, although the problem of the vanishing cosmological constant was a nagging one. Since decades back it had been thought to be zero, and there were unsuccessful attempts to prove this fact using fundamental physics. 

This all changed towards the end of the millennium. The turning point was the shocking discovery of a non-vanishing cosmological constant (cc) \cite{Riess:1998cb, Perlmutter:1998np}. This observation implied we were in desperate need of a principle forcing the cc to more or less vanish on the scales of particle physics, there also had to exist further subtle effects allowing for a tiny non-zero value. The required amount of fine-tuning was completely unnatural from the point of view of string theory and seemed impossible to account for. The solution many finally adopted had actually been proposed ten years earlier by Barrow \cite{Barrow}, Barrow and Tipler \cite{Barrow:1988yia} and Steven Weinberg \cite{Weinberg:1988cp}. They made the phenomenological observation that there was a limit on how big a hypothetical cosmological constant could be and still allow for a universe where galaxies could form. Since there is no reason for the cosmological constant to be much smaller than this, they argued that a value a bit lower than the maximal one would be the most natural. Remarkably, this is more or less the value of the cosmological constant we observe. 

One can argue about the details of this argument, but give or take a couple of orders of magnitude, it was spot on given the rules of the game. What was needed in order for the argument to make sense was the {\it anthropic principle} \cite{CARTER}. In simple terms it is no more than the assumption that the world is much bigger and more varied than we so far have been able to explore. In particular, this applies to the value of the cosmological constant, which is assumed to vary randomly from one region to another, with its seemingly fine-tuned value as just a necessary property of a local, habitable environment. Other fine-tunings required for the existence of stars, and ultimately life, is explained in the same way.  This leads to the notion of a {\it multiverse} of which our universe is just a tiny part. Whether and how you can reach other regions of the multiverse is not essential for the argument, what is important is that they do exist in the same way as ours.

The anthropic principle was embraced by some relieved string theorists \cite{Bousso:2000xa, Susskind:2003kw}, whose work now became so much easier. The idea of a landscape of possible string vacua was born, that laid the foundations of a multiverse in string theory \cite{Schellekens:2015zua}. The preconception that such a landscape simply had to exist lowered the bar of mathematical rigor. The difficulty of the subject makes it impossible to take all effects that contribute to vacuum energy into account and you are forced to make simplifications that, hopefully, can be justified when you see the end result. If, given certain simplifications, you find a result that is in line with the general expectations this can be viewed as justification enough. In this way the landscape was quickly populated \cite{Kachru:2003aw} and it was argued to provide a possible explanation of the cosmological hierarchy problem \cite{Susskind:2003kw, Tye:2018ist}.

Luckily, other parts of cosmology developed at a quick pace lending growing support for the theory of inflation. A prerequisite for inflation, is the existence of a potential energy that drives the accelerated expansion of the early universe in a way very similar to our late time dark energy. This potential energy is in turn determined by the value of a scalar field. Remarkably, calculations show that quantum fluctuations of the value of such a field lead to a CMB spectrum of just like what we observe. Interestingly, scalar fields of this very kind abound in string theory and are an integral part of the string landscape. Possible states of the universe correspond to minima in a multidimensional moduli space spanned by the numerous different scalar fields. A picture emerges, where the local state of the universe can shift from one minimum to another. Inflation is just the process where our universe came to rest at one such particular minimum, while we expect the outcome to be a different one at other places. In this way, inflation can be viewed as our first glimpse of the multiverse.

These ideas became the target of string theories most fierce critics. The claim that the landscape includes of the order of $10^{500}$ different vacua became almost a joke. How could a theory with that amount of freedom have any chance of predicting anything? This kind of criticism is, however, misguided \cite{Dawid:2013maa, Polchinski:2015pzt}. One might compare with quantum field theory, where there is an infinity of fully consistent theories. Experiments are needed to pick the right one, and parameters must be fitted. When this is done the theory still has enormous predictive power, and no one would claim that the Standard Model is useless. One could argue in a similar way concerning the string landscape. Measurements need to be made in order to figure out where in the landscape you sit, including the size and shape of the extra dimensions, and only when this is achieved one can hope to make non-trivial predictions. Possible ways to test the theory includes high precision measurements in particle physics, and observations of the early universe \cite{Quevedo:2016tbh,  Silverstein:2017zfk}.

The problem with the string landscape that we want to address, does not concern whether the multiverse is a reasonable proposal or not. We remain agnostic about its size and character, and about which quantities have an anthropic explanation. Historically, one can argue that the idea of a multiverse is a rather conservative approach with many parallels, and that attempts to portray it as a useless speculation are based on misunderstandings. The real problem is a completely different one. Paradoxically the critics of string theory and the proponents of the string landscape all agree on one thing: the landscape exists and we more or less know its properties. But what if they are wrong?     
    
During the last almost two decades, much work on string cosmology has been the victim of a circular reasoning:  we observe de Sitter space, we like string theory, hence we conclude that string theory must contain de Sitter space. Pointing towards the astronomical observations of dark energy, preliminary calculations purporting to show the existence of de Sitter vacua are accepted to be essentially correct. In turn, such calculations are used to argue that string theory makes successful predictions thus completing the circle. As we will argue, it is far from established that there exists a landscape of de Sitter vacua in string theory. On the contrary, there is mounting evidence that string theory abhors de Sitter space with unexpected conspiracies showing up as soon as you are in a position to actual perform detailed calculations. As we will review, \emph{there is not a single rigorous 4D de Sitter vacuum in string theory}, let alone $10^{500}$. All claimed examples rely on assumptions about the outcome of calculations no one has yet been able to perform. In line with the overarching paradigm it is standard procedure to claim that it is all just a matter of unimportant technical details without any baring on the overall conclusion. 

We feel that it is time to change attitude. String theory has matured far enough to stand on its own, we need to explore the theory with an open mind, acutely paying attention to what it actually has to say about the existence of de Sitter vacua and dark energy. The possibility still remains that string theory fails to provide a single de Sitter vacuum \cite{Brennan:2017rbf}, and thus could miss out on its first observational test. It is also possible that string theory is the correct theory of quantum gravity, but the way it is made compatible with observations is much more sophisticated than what we now are attempting.

The rest of this paper is organised as follows. In  section \ref{sec:constructions} we review various proposed de Sitter vacua in string theory, which are then criticized in section \ref{sec:problems}. In section \ref{sec:nodS} we discuss the possibility that there are no de Sitter vacua at all, and what it implies and we conclude briefly in section \ref{sec:conclusion}.
\section{dS constructions in string sofar} \label{sec:constructions}

\subsection{Conceptual framework}
Before we discuss actual constructions of vacuum solutions in string theory it is important to remind ourselves of the conceptual framework of flux compactifications that allows us to compute vacuum energies. 

It is standard to construct vacuum solutions, with stabilised moduli, in weakly coupled string theory in the form of 10D supergravity with some leading-order corrections. A minimal set of criteria that guarantees consistency are
\begin{itemize}
\item
The string coupling $g_s$ is stabilised at small values, $g_s<<1$.
\item All fields have low gradients in order to suppress derivative corrections. 
\item Extra dimensions have to be large enough in order to ignore winding states (and again derivative corrections to the effective action).
\end{itemize}

If one further insists on having a 4D effective description one requires a parametric separation between the size of the extra dimensions $L_{KK}$ and the size of the observable universe $L_{\Lambda}$. Note that this is automatically satisfied if the vacuum is Minkowski, but in case of AdS or dS it requires that the cosmological constant is small in $M_{KK}$ units. 

The cc is then claimed to be the minimum of the effective potential $V(\phi)$ for the moduli $\phi^i$ that appears in the effective action:
\be
S_{eff}=\int\sqrt{|g|}\Bigl(R -\tfrac{1}{2}G_{ij}\partial\phi^i\partial\phi^j -V(\phi)\Bigl)\,.
\ee
But this begs the question of whether one is computing the bare cc or the actual cc?  Naively one could expect that we are computing the bare cc since the potential is a classical potential in 4D, although it incorporates corrections to 10D SUGRA. Actually, following the rules we have set up, the bare cc will be close to the full vacuum energy, even though it would seem unnatural from a QFT point of view, where the quantum contributions are highly sensitive to the UV degrees of freedom, and the natural values for the cc would be of the order of the cut-off scale of the effective field theory. However, we are working in a fine-tuned corner of the moduli space, where the weakly coupled space-time solution is argued to be a good approximation to the full string theory solution. Hence, the computed cc is close to the full answer and there are no extra quantum contributions changing its value. This is why this framework of self-consistent flux compactifications can actually compute true vacuum energies, which is quite a remarkable achievement from a pure field theory viewpoint. 

In fact the standard lore that matter multiplets backreact on the vacuum energy (by means of loop corrections) is still true and visible in the string theory constructions. Fore instance, imagine having a controlled moduli-stabilisation mechanism that relies on fluxes and leading corrections to the 10D action, such that one ends up with a self-consistent meta-stable dS vacuum. Now imagine trying to ``add the standard model" by means of  extra branes that wrap certain cycles. This certainly upsets the moduli-stabilisation and one needs to redo the computation, check the new RR tadpole conditions, and compute the backreaction on the moduli. Only in special circumstances can one find a new dS vacuum that is close to the original one. This shows  that in string theory, particle physics and cosmology are inseparable, which is ultimately what the cc problem is telling us: the vacuum energy (cosmology) is completely dependent on the details of the UV physics (particle physics).\footnote{It is likely that the above viewpoint is not shared by all researchers working on this. For instance in \cite{Balasubramanian:2004uy} one can find a discussion about 1-loop corrections to cosmological constant computed in the KKLT scenario that is argued to be large. This also relates to the conceptual difference between the Bousso-Polchinski (BP) scenario \cite{Bousso:2000xa} and the KKLT-type scenarios. In BP there is room for a bare cc piece that can come from matter loops. What matters in BP is that on top of this you can add a dense energy spectrum from the fluxes that can cancel the bare cc.}

In what follows we overview the dS constructions known in string theory. Whereas in section \ref{sec:problems} we discuss the possible issues, assuming at least the conceptual framework is correct. One should keep in mind that there is still a possibility that the above framework is inconsistent from the start and one cannot  construct effective field theories from string theory around vacua that are not the original SUSY backgrounds one uses to quantise the string (10D Minkowski, or 4D Minkowski$\times$CY$_3$,...). See \cite{Banks:2012hx} for some remarks on this issue. 

The reader interested in understanding the details of flux compactifications can consult the reviews \cite{Silverstein:2004id, Grana:2005jc, Douglas:2006es, Denef:2008wq,Denef:2007pq, Samtleben:2008pe, Ibanez:2012zz, Baumann:2014nda} .
\subsection{Classification scheme}

In what follows we present a (very) rough sketch of the known methods to construct de Sitter vacua in string theory. Unfortunately there is no review paper on this topic\footnote{An interesting light-version of a review on dark energy in string theory can be found in \cite{Greene:2015fva}} that contains all the relevant references. As there are several hundred papers on the topic we will not attempt to review them all. Instead we present a biased set of references that are selected on the basis of the memory of the authors and what we think were original references, in the sense of being amongst the first to propose a mechanism for dS model building. The number of papers published on this topic strongly depends on ones definition of a dS construction in string theory. Quite some papers are not genuinely top down and often discuss dS vacua in `stringy-inspired set-ups'. In this paper we aim to review specifically the most top-down constructions.

To bring structure in the jungle of papers on dS constructions, we attempt the following classification scheme. First we can divide the constructions according to the compactness of the extra dimensions. As is well-known since the celebrated Randall-Sundrum paper \cite{Randall:1999vf}, extra dimensions can be non-compact if the warping is sufficiently strong and if the gauge theory is somehow confined to live on a 4-dimensional hypersurface (D-brane). Some attempts to find dS solutions in these brane world scenarios can be found in \cite{Gibbons:2001wy, Neupane:2010ey, Minamitsuji:2011gp}. In what follows we will not discuss this and simply restrict to compact models. 

Second, we divide the constructions into critical versus non-critical string theory. Historically the first attempts to construct dS solutions were in fact made in non-critical string theory \cite{Silverstein:2001xn, Maloney:2002rr}\footnote{See also \cite{Dodelson:2013iba}.}. The main idea here is that the effective action (in string frame) now gets an extra contribution:
\be
S = \tfrac{1}{2\kappa^2}\int\sqrt{g}e^{-2\phi}\Bigl( R - \frac{2(D-D_c)}{3}M_s^2 +\ldots \Bigr)\,.
\ee
with $M_s$ the string mass scale and $D_c$ the critical dimension. So when $D>D_c$ there is an extra positive contribution to the effective potential, which seems to be rather useful in trying to find de Sitter solutions after compactification. As the authors of   \cite{Silverstein:2001xn, Maloney:2002rr} mention it seems difficult to understand whether there is truly any perturbative control over these models, so for that reason we will not discuss them any further, although the idea of going beyond criticality should probably be looked at more.

\begin{figure}[h!]
\centering
\includegraphics[width=.9\textwidth]{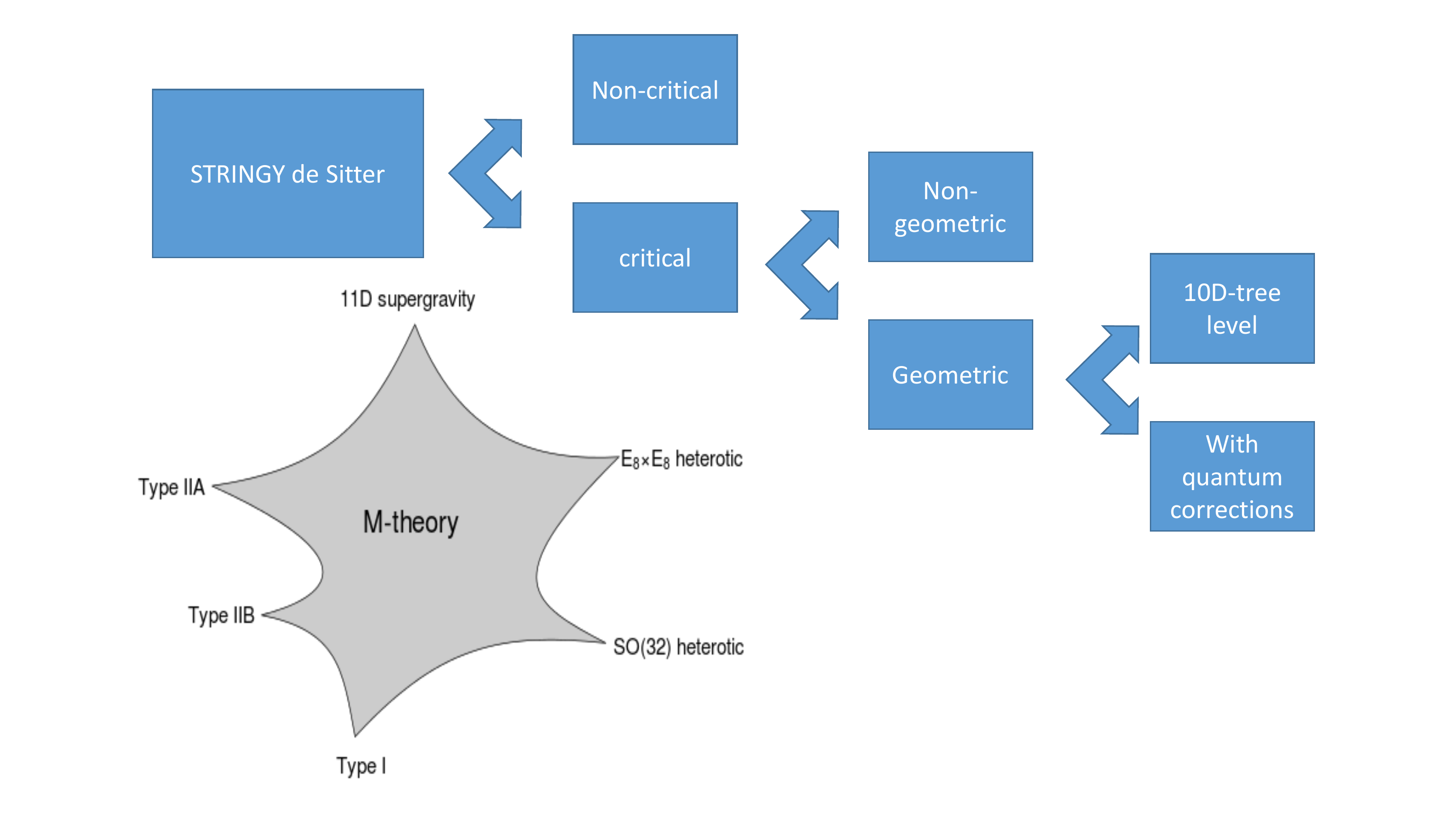}
\caption{\small{An attempt to classify de Sitter constructions in string theory }}%
\label{Figure1}%
\end{figure}

The critical models can be further divided into geometric versus non-geometric constructions. The geometric constructions can further be divided into ``classical'' versus ``quantum'' solutions. For each of the branches of critical string theory constructions one can further classify depending on where the model is situated in the web of the 5-string theories and 11-dimensional supergravity (M-theory). This classification is pictorially presented in figure \ref{Figure1}.

\subsection{Classical}
Perhaps the most obvious manner to construct de Sitter vacua uses 10-dimensional supergravity with fluxes at the two-derivative level. This means we start with a 10D action, which is of the schematic form:
\be\label{action}
S=\int \sqrt{|g|}\Bigl(R -\tfrac{1}{2}(\partial\phi)^2 - \sum_p \tfrac{1}{2(p!)}e^{a_p\phi}F_p^2 \Bigr) + \text{Chern-Simons}\,,
\ee
where $F_p$ are the rank-p fieldstrengths of 10-dimensional supergravity (for instance odd $p$ in IIB). The scalar potential $V$ that can arise from a compactification then comprises two kinds of terms:
\begin{equation}
V =  V_f + \sum_p V_p\,,
\end{equation}
where $V_f$ arises from integrating over the curvature of the extra dimensions:
\begin{equation}
V_f = - \int\sqrt{g_6}R_6\,,
\end{equation}
and $V_p$ is the dimensional reduction of the flux terms
\be
V_p = \int\sqrt{g_6}\tfrac{1}{2(p!)}e^{a_p\phi}F_p^2 \,.
\ee
Hence we see that $V_f$ is positive (negative) if the internal dimensions are curved negatively (positively), whereas the flux contributions are always positive\footnote{In case a flux is space-time filling $F^2$ is negative, but one can show that one generates an extra minus sign in the potential. Hence it is preferable to work in the democratic formalism where one can avoid this by writing the Hodge dual flux.}.

Well-known nogo-theorems, such as \cite{Maldacena:2000mw}, imply that this framework cannot give dS vacua if the internal space is static, compact and without singularities. The static condition cannot be broken in any simple way, since vacuum solutions have their moduli stabilised. The compactness might be dropped if the warping is significant enough to create a gapped KK spectrum, as we argued earlier. But the most conservative way out relies on the use of physical singularities, ie localised brane sources, that can be added to the 10D action. This generates extra terms in the 4D scalar potential of the form
\be
V_{\text{source}} = \mu\int\sqrt{|g_n|}\,,
\ee
where $\mu$ is the tension of the brane and $g_n$ the determinant of the 10D metric pulled back on the internal submanifold wrapped by the brane.

It is not difficult to show that ordinary D-branes do not help in finding dS vacua but negative tension sources, that is orientifold planes, are crucial \cite{Hertzberg:2007wc, Silverstein:2007ac, Danielsson:2009ff, Wrase:2010ew} (see also \cite{Dasgupta:2014pma}) for further nogo-theorems with those ingredients). This is not only true for dS spaces, but also for Minkowski vacua (with fluxes) and even AdS vacua if one insists on the KK scale to be parametrically smaller than the AdS length scale \cite{Gautason:2015tig}. 

String theory vacua, build from these ingredients, are conceptually easy to understand. The solution arises as a consequence of a balance of forces: the negatively curved dimensions want to dynamically expand to lower their energy, cycles threated by fluxes also want to expand to lower the flux densities, but then the presence of the sources counteracts these runaways. 

Not a single dS solution without a tachyon has ever been found along these lines and it remains an excellent problem to construct one\footnote{Although some partial success was claimed for 3D solutions that involve non-geometric branes \cite{Dong:2010pm}. Non-geometry is discussed further below.} or prove it cannot be done \cite{Chen:2011ac, Danielsson:2012et, Junghans:2016abx, Junghans:2016uvg, Andriot:2016xvq, Andriot:2017jhf}.

Most of the work on constructing dS critical points in this setup was carried out in IIA supergravity with O6 sources, see for instance \cite{Caviezel:2008tf, Flauger:2008ad} for  the first solutions ever constructed  and \cite{Danielsson:2011au} for a review and a large scan of solutions. All these solutions necessarily feature non-zero Romans mass and negatively curved compact space. The O6 configuration tends to be rather involved since it seems that intersecting orientifolds are crucial \cite{Andriot:2016xvq, Andriot:2017jhf}. The known examples have O6/D6 intersections of the following kind:
\begin{center}
  \begin{tabular}{|c|c|c|c|c|c|}
    \hline
    \rule[1em]{0pt}{0pt} $\bigotimes$ & $\bigotimes$  & $\bigotimes$ & -- & -- & -- \\\hline
    \rule[1em]{0pt}{0pt} $\bigotimes$ & -- & -- & $\bigotimes$  & $\bigotimes$ & -- \\\hline
    \rule[1em]{0pt}{0pt} -- & $\bigotimes$ & -- & -- & $\bigotimes$  & $\bigotimes$ \\\hline
    \rule[1em]{0pt}{0pt} -- & -- & $\bigotimes$ & $\bigotimes$ & -- & $\bigotimes$ \\\hline
   \end{tabular}
\end{center}
where a $\bigotimes$ symbolises a direction of a O6/D6 source.

An interesting and not well understood property of these solutions is their scarcity. The typical orientifold compactification with fluxes and negatively curved extra dimensions features various critical points of the scalar potential if sufficient fluxes are turned on. But they are almost always non-SUSY AdS vacua. Typically one requires a very high amount of fine-tuning to find dS critical points and they are consequently far outnumbered by the AdS solutions, see for instance the plots in \cite{Danielsson:2010bc}.

\subsection{Non-geometric}
It is by now well-known \cite{Shelton:2005cf} that a T-duality chain, starting along a circle threated by NSNS 3-form flux $H_3$ leads to ``metric-flux" and subsequently to non-geometric fluxes, which have no 10D origin that can be found in the action (\ref{action}). Let us follow the notation of \cite{Hertzberg:2007wc} and write the metric in 10D string frame as follows
\be
\d s^2 = \tau^{-2} \d s^2_4 + \rho \d s_6^2\,,
\ee
where we ignored the warpfactor because we work in the approximation that the sources are smeared over the extra dimensions. We have explicitly shown the dependence on two universally-present scalars: $\rho$ and $\tau$. Since $\d s^2_6$ has fixed volume (say $1$ in string units), the scalar $\rho$ measures the volume and hence the KK scale. The scalar $\tau$ is given
\be
\tau^2 =\exp(-2\phi)\rho^3\,,
\ee
and is required in order for $\d s^2_4$ to be in Einstein frame. All ingredients to the scalar potential have universal scalings w.r.t.~$\tau$ and $\rho$. For instance the NSNS 3-form flux $V_3$ and the curvature $V_f$ go like
\be
V_3\sim \rho^{-3}\tau^{-2}\,,\qquad V_f\sim \rho^{-1}\tau^{-2} \,.
\ee
The essence of non-geometric fluxes $Q$ and $R$ is that they come with the following scalings \cite{Hertzberg:2007wc}:
\be
V_Q\sim \rho\tau^{-2}\,,\qquad V_R\sim \rho^{3}\tau^{-2}\,,
\ee
which cannot be obtained from the 10D action and turn out very useful in dS model building \cite{Hertzberg:2007wc}.  At the level of 4D sugra it is not difficult to implement these non-geometric fluxes. Within, say, the context of minimal supergravity it is rather straightforward to add terms to the superpotential that follow from the T-duality chain \cite{Shelton:2005cf}. This can be done also at the level of extended supergravity, and has provided many of the less understood gauged supergravity theories with a ``stringy" origin. 

In a given duality frame there exist a number of fluxes that can be given a geometric interpretation. This include RR-fluxes, H-fields, and metric fluxes. Other fluxes can not be interpreted in this way. When the duality frame is changed, for instance by performing  a T-duality, which connects type IIA and type IIB, geometric and non-geometric fluxes can be interchanged. For a given flux it seems preferable to choose, if possible, a frame in which it has a geometric interpretation. For some combinations of fluxes such a frame does not exist \cite{Dibitetto:2012rk}, and the background is inherently non-geometric\footnote{From the point of view of supergravity it is natural to extend the allowed fluxes to include not only the geometric ones, but also the non-geometric. In simple type IIB toroidal orientifolds the number of `isotropic' geometric fluxes is 8, while the total number of geometric fluxes is 16. In type IIA the isotropic cases again involves 8 fluxes, but the total number is  24. The non-geometric generalization increases the numbers to 32 and 128, respectively.}

In this extended framework, with an increased number of free parameters, it turns out to be possible to find metastable dS-vacua. They are scarce, but they exist. Using the $\mathcal{N}=1$ superpotential of \cite{Shelton:2005cf} the very first tachyon-free de Sitter vacua were constructed \cite{deCarlos:2009fq, deCarlos:2009qm}. These models contained only a handful of complex scalars, and one could worry whether tachyons appear once certain twisted moduli were considered. Interestingly stable solutions seem to continue to exist even when more scalars are added \cite{Danielsson:2012by, Blaback:2013ht} (see also \cite{Damian:2013dq, Damian:2013dwa, Blumenhagen:2015xpa}). Nonetheless the scarcity that was seen in the geometric case still persists. Typically the dS vacua hide in very small  corners in the landscape of allowed fluxes.

\subsection{Quantum}
The overwhelming majority of de Sitter constructions in the literature are based on classical flux compactifications with orientifold sources \cite{Dasgupta:1999ss,Giddings:2001yu} but add in ``quantum effects'' in order to stabilise all the moduli. By ``quantum effects'' we mean everything beyond 10-dimensional supergravity action at the two-derivative level (including brane sources).  Such corrections comprise higher-derivative corrections, string loop corrections, non-perturbative corrections in the string coupling, etc. Clearly this makes those constructions more involved as the exact computation of such corrections is hard. 

There are two main motivations for resorting to this difficult corner. On the one hand, the existing de Sitter nogo-theorems suggest that it is already difficult (but not impossible) to find dS at the classical level. Secondly, regardless of the sign of the cosmological constant, quantum corrections are typically necessary to stabilise all moduli. 

It is well-known that the nogo-theorems based on classical 10D sugra \cite{Maldacena:2000mw} can be extended to incorporate some quantum effects.  Most work on de Sitter nogo's beyond the 10D supergravity (at two-derivative level) has been carried out for heterotic string theory \emph{at string-tree level}, where the following observations have been made (in chronological order)
\begin{enumerate}
\item  The classical supergravity equations of motion show that the sub-leading $\alpha'$ corrections cannot give rise to dS vacua \cite{Green:2011cn}. 
\item This argument was then simplified and extended to the infinite tower of $\alpha$' corrections in \cite{Gautason:2012tb} where the authors further remarked that even AdS vacua are forbidden.
\item  One could then still wonder whether these results are due to the perturbative approach and maybe non-perturbatively stringy derivative corrections can help. These hopes dissipated after it was shown that a worldsheet analysis is possible implying no  dS solutions \cite{Kutasov:2015eba}. 
\item The most recent analysis \cite{Quigley:2015jia} has pushed the techniques of \cite{Gautason:2012tb} to include even the effects of gaugino condensates and showed that again dS (and AdS) vacua are excluded.
\end{enumerate}
The above nogo theorems apply to the heterotic string, but it is natural to expect that dualities carry this over to other string theories, implying similar difficulties in achieving dS vacua through quantum constructions, although all the dS constructions discussed in the next sections go around the assumptions of the mentioned nogo theorems. Nonetheless, the existing nogo theorems at least show that quantum effects are not the magical ingredients that make dS building work. 

\subsubsection{IIB string theory}

Let us focus the discussion on the best-understood flux models: IIB string theory with 3-form fluxes and sources carrying 3-brane charges (D3/O3, D7/O7), which were pioneered in \cite{Dasgupta:1999ss, Giddings:2001yu}. These constructions naturally lead to Minkowski vacua at tree-level where a specific set of moduli remains unstabilised by the fluxes. The moduli are then further stabilised when quantum corrections are considered. 

Whenever 3-form fluxes are present, the $F_5$ Bianchi identity (equation of motion) leads to the following topological constraint:
\begin{equation}\label{tadpole}
Q_3 = \int H_3\wedge F_3\,,
\end{equation}
which is often referred to as the RR tadpole condition and will play an important role in what follows. Here $Q_3$ is the total amount of 3-brane charges carried by all the sources in the compactification. To derive the Minkowski solutions we will ignore all warping effects and the $F_5$ field that is generated by the sources. The energy contained in the warping and the $F_5$ are accounted for by the DBI term. So let us consider the IIB action and isolate the three-form fluxes and the (smeared out) D3/O3 DBI term. The contribution to the scalar potential in four dimensions is
\begin{equation}\label{eq:smearpot}
V = \int_6 \sqrt{g_6}\Bigl(\frac{1}{2(3!)}g_s^{-2} H_3^2 + \frac{1}{2(3!)}F_3^2 + \frac{\mu}{Vol_6}\Bigr) \,,
\end{equation}
where we \emph{assumed} that the internal geometry is Ricci flat. If we can find a critical point of the scalar potential, then this assumption was self-consistent. The tadpole constraint, together with the demand that the brane-setup is BPS, allows us to rewrite the DBI contribution:
\begin{equation}
 \mu = \mp g_s^{-1} Q  = \mp g_s^{-1} \int H_3\wedge F_3\,.
\end{equation}
This implies that the potential becomes a total square:
\begin{equation}\label{square}
V = \frac{1}{2}\int_6 \Bigl\{g_s^{-1}\star H \pm F_3\Bigr\}^2\,.
\end{equation}
The $\pm$ choice depends on whether the source has charge equal to plus or minus its tension. This can be chosen at will and  can be seen as  a convention. Since the potential is a sum of squares we know that a vacuum solution exists whenever the squares are zero:
\begin{equation}\label{ISD}
g_s^{-1}\star H \pm F_3 = 0\,.
\end{equation}
In this paper we take the plus sign convention. This is referred to as the ISD condition, where ISD stands for Imaginary Self Dual, because the complex three form $G = F - i g_s^{-1}H_3$  obeys $\star_6 G = iG$. The source tension is now necessarily negative since
\begin{equation}
\mu = - \int  F_3^2 \,.
\end{equation} 
So there is a net negative tension from orientifolds. The Minkowski vacuum arises from an exact cancellation between the positive energy of the fluxes and the negative tension energy in the sources. 
These solutions preserve supersymmetry if the internal manifold has maximally $SU(3)$ holonomy (Calabi-Yau) and the $G_3$ flux has only primitive $(2,1)$ pieces.

Clearly the Minkowski vacuum is a minimum of the potential and the masses of the various scalars are therefore either zero or positive. It turns out that typically one can stabilise all complex structure moduli of the CY space but the K\"ahler moduli remain unfixed. The latter can be stabilised by the quantum effects. 

Before we can introduce the quantum effects we need a few more details about the 4D supergravity description. The scalar potential can be deduced from the following superpotential\cite{Gukov:1999ya}
\be \label{GVW}
W = \int_6 \Omega_3\wedge G_3\,,
\ee
where $\Omega_3$ is the holomorphic 3-form of the CY space. The form $\Omega_3$ carries all the dependence on the complex structure moduli, whereas the K\"ahler moduli only appear in the potential via the K\"ahler potential $K$ (which also depends on the complex structure) via the standard formula
\be
V =  e^K\Bigl(K^{a\bar{b}}D_aWD_{\bar{b}}W -3 W^2\Bigr)\,,
\ee
with $D_a W = \partial_a W-  W\partial_a K$.  The indices $a, b$ run over all moduli. In what follows  the indices $\alpha,\beta$ run over the K\"ahler moduli and the indices $i, j$ over the complex structure moduli. One can show that the K\"ahler potential obeys $K^{\alpha\bar{\beta}}\partial_{\alpha}K\partial_{\bar{\beta}}K =3$, such that the potential becomes a sum of squares:
\be
V =  e^K\Bigl(K^{i\bar{j}}D_iWD_{\bar{j}}W\Bigr)\,,
\ee
which is nothing but the earlier sum of squares (\ref{square}). This potential is called of the no-scale type. It still depends on the K\"ahler moduli through the dependence in $K$, but not inside the vacuum since then $D_i W=0$ and the mass terms for the K\"ahler moduli vanish. 

The essence of the KKLT scenario \cite{Kachru:2003aw} is the use of quantum corrections to the superpotential, which creates a dependence on the K\"ahler moduli. Well-known non-renormalisation theorems imply that the only corrections to $W$
must be of a non-perturbative nature. KKLT considered for simplicity the case of single K\"ahler CY spaces, whose complex K\"ahler modulus we  denote $T$. The non-perturbative corrections can come from either Euclidean D3 instantons that wrap a supersymmetric 4-cycle or from ``fractional instantons'' that come from a stack of 7-branes that wrap this four-cycle. This stack of 7-branes has a $N=1$ SU(N) SYM theory living on it which undergoes gaugino condensation in the IR. This means that the gauge theory has a (leading) superpotential term of the form
\be
W = \Lambda^3 \exp(\frac{2\pi i}{N} T)\,,
\ee
where $\Lambda$ is dynamical scale of the gauge theory. The modulus $T$ appears since it is the effective coupling of the gauge theory. Let us go back to the bulk theory with gravity and look at the case when supersymmetry is broken by a $(0,3)$ piece in $G_3$, such that the on-shell value of $W_0$ from the fluxes (\ref{GVW}) does not vanish. Now the assumption is that the superpotential induced by the gauge theory can simply be added to the flux-induced superpotential, such that the result is
\be
W(T) = W_0 + \Lambda^3 \exp(\frac{2\pi i}{N} T)\,,
\ee
where $W_0$ is now a constant, since we assume the complex structure moduli are integrated out as they have high masses (order string scale). The resulting scalar potential for the imaginary part of $T$ has a supersymmetric AdS minimum, which is sharply peaked and can be at very small values for the energy, if some fine-tuning for $W_0$ is possible. Note that the imaginary part of $T$ is a function of the volume of the internal space, so this procedure stabilises the overall volume. Self-consistency implies that this volume must be parametrically larger than the string scale, otherwise derivative corrections might spoil the consistency. At the same time this guarantees that higher instanton corrections are consistently sub-dominant.

It is of importance to pause here before we get to de Sitter. What we described sofar is quite non-trivial: this procedure achieves a vacuum with full moduli-stabilisation and at the same time a parametric difference  between the length-scale of the observable dimensions (the AdS length) and the non-observable ones  (the KK-scale)
\be
\frac{L_{KK}}{L_{AdS}}<<1\,.
\ee
This means that the vacuum is perceived as genuinely four-dimensional to an observer at low energies. This is a striking achievement since prior to this construction it was unclear whether string theory had any calculable vacua satisfying the two most important phenomenological constraints of moduli-stabilisation and scale separation. Note that Freund-Rubin solutions like $AdS_4 \times S^7$ (in 11D supergravity) have moduli-stabilisation but not scale separation, such that they cannot be seen as 4D vacua and they are hence useless to phenomenology (but highly valuable to holography). 

KKLT went further and suggested an explicit way to turn this supersymmetric AdS vacuum into a de Sitter vacuum, by adding anti-D3 branes \cite{Kachru:2003aw}. Anti-D3 branes, in this context, mean D3 branes with a charge density that has the opposite orientation to the six-form $H_3\wedge F_3$. In contrast with D3 branes, anti-D3 branes break supersymmetry of the background and add positive energy
\be \label{uplift}
V_{\bar{D3}} = 2 \mu \exp(4A)\,,
\ee 
where $\exp(A)$ is the warpfactor in front of the 4D metric
$\d s_{10}^2 = \exp(2A)\d s_4^2 + \d s^2_6$ evaluated at the position of the $\overline{D3}$. The warpfactor dependence will turn out crucial as explained below. The uplift energy (\ref{uplift}) equals twice the DBI action of the $\overline{D3}$-brane. The factor of 2 can be traced back to the fact that the $\overline{D3}$ leads to two sources for positive energy: the tension of the anti-brane itself, equal to $\mu \exp(4A)$, and an increase in 3-form fluxes to cancel the tadpole. Also this energy is equal to $\mu \exp(4A)$, which can be derived along the previous reasonings we made\footnote{Alternatively one can imagine adding in a $D3/\overline{D3}$ pair \cite{Maldacena:2001pb}. This does not affect the 3-form fluxes, but gives twice the tension. But the D3 itself does not add energy, hence the $\overline{D3}$ must add twice the energy.}. 

The claim of KKLT is now that this lifts the AdS vacuum to a meta-stable dS vacuum if the $\overline{D3}$ tension term is tuned properly. This rests on three observations. 
\begin{enumerate} 
\item The dependence of the $\overline{D3}$ tension on the K\"ahler modulus is polynomial, whereas the instanton-generated AdS potential was sharply peaked due to the exponential dependence. Hence the anti-brane tension term is almost constant for small excursions around the AdS minimum, such that it effectively adds a constant. In particular it does not create a tachyonic direction in the K\"ahler or complex structure moduli. 
\item The above can only be consistent when the uplift term is not too big, otherwise the local minimum is not only lost, but also the compactification will not be under control due to backreaction effects. Luckily, one can argue that warped Calabi-Yau solutions with three-form fluxes have local throat regions. Such regions can for instance look like the well-known Klebanov-Strassler throat solution \cite{Klebanov:2000hb}. The $\overline{D3}$ feels a force towards the minimum of the throat where $\exp(A)$ is minimized. This implies that the typical string scale energy of a $\overline{D3}$ will be dynamically warped down to tunable small values. Hence the supersymmetry breaking ingredient is argued to be tunably small, which is required for a successful uplift. 
\item  The $\overline{D3}$ branes, despite breaking supersymmetry and carrying charges opposite to the fluxes,  will not annihilate immediately with the background fluxes. When the $\overline{D3}$ charge is sufficiently small one can use a probe argument to show there is a classical barrier against brane-flux annihilation \cite{Kachru:2002gs}. This means that also the open-string sector is (meta-)stable. 
\end{enumerate}
This establishes the KKLT dS minimum. It is however far from a unique construction in this setup. Let us mention some ideas that can be considered as variations upon the same conceptual theme. 

First of all there are different ways to stabilise the moduli of the no-scale Minkowski solutions. Most notably there is the Large Volume Scenario (LVS) \cite{Balasubramanian:2005zx}. The LVS scenario uses that there is a non-SUSY AdS vacuum that can be found when considering the leading $\alpha'$ correction to the K\"ahler potential. It is believed that this vacuum is self-consistent and safe from further  corrections to the effective action. The virtue of this vacuum, with respect to the KKLT AdS vacuum, is that the volume is typically much larger pushing the solution into a regime where corrections are under better control. To turn the LVS solution into a dS vacuum one can attempt anti-brane uplifting, in case warped throats are available. 

Instead of considering different corners of string theory with stabilised moduli one can consider different methods to uplift AdS vacua. For instance, instead of the energy carried by $\overline{D3}$-branes, one can consider energy from SUSY-breaking in a hidden sector \cite{Lebedev:2006qq}, which would effectively be an F-term uplifting. There is also a notion of D-term uplifting \cite{Burgess:2003ic, Parameswaran:2006jh, Achucarro:2006zf} with D-terms generated by magnetised 7-branes but this can be subtle \cite{Choi:2005ge, Villadoro:2005yq}.

An alternative method to uplift uses the AISD components of the 3-form fluxes since these would add positive energy and have a similar powerlaw dependence on the volume modulus \cite{Saltman:2004sn}. For this to work there have to be local minima of the complex structure moduli which are not making the fluxes ISD. This was argued to be possible \cite{Saltman:2004sn}. Morally this is not too different from anti-brane uplifting since one can think of AISD fluxes as dissolved anti-branes. However one cannot rely on warped throats to redshift the uplift energy, instead and one can hope that the large flux numbers can make the non-ISD minima parametrically small, in order that a successful uplift can be achieved. We refer to \cite{Gallego:2017dvd} for the state-of-the-art of flux-induced F-term uplifting.

In the last 10 years many more de Sitter constructions in IIB have appeared and it is difficult to do justice to all. Let us simply summarize them as follows: these constructions can be seen as moduli-stabilisation scenario's based on the GKP solutions. By constructing an effective potential as the sum of the tree-level GKP piece and all other corrections (string loops, instantons, gaugino condensation, $\alpha'$-corrections) these papers claim to achieve self-consistent vacua. This means that these vacua are on specific positions in moduli space for which one can argue that all other, ignored, corrections are subleading. See \cite{Westphal:2006tn, Parameswaran:2007kf, Rummel:2011cd, Louis:2012nb, Cicoli:2012fh,Cicoli:2013cha, Aparicio:2014wxa, Cicoli:2015ylx, Antoniadis:2018hqy} for an incomplete list. General conditions on the perturbative stability for such solutions, implied by the sgoldstino, were analysed in \cite{Covi:2008ea, Covi:2008zu} and provide interesting constraints on model building.

\subsubsection{Other string theories}
Surprisingly much less research went into the study of dS vacua from corrections to the 10D SUGRA actions of string theories different from IIB. One would expect that the natural starting place are the classical Minkowski vacua from IIA orientifolds that one can engineer by T-dualising the IIB orientifolds\footnote{Of course only simple toroidal orientifolds can be T-dualised explicitly. What we mean is the construction of IIA orientifolds with the T-dual ingredients.} \cite{Kachru:2002sk, Grana:2006kf, Grana:2005jc,Blaback:2010sj, Andriot:2016ufg}. Then one could for instance mimic the KKLT/LVS constructions, but using IIA ingredients. To our knowledge this has been tried only once \cite{Palti:2008mg}.

Instead some investigations have started directly with quantum corrections to the effective action, assuming some classical flux mechanism would stabilise many other fields in an AdS vacuum and observe that the right kind of instantons exist to possibly ``uplift' the AdS vacua \cite{Davidse:2005ef, Saueressig:2005es}. 

We already mentioned that IIA string theory is a more natural environment to study moduli-stabilisation, since it is suggested that moduli-stabilisation can even be achieved at the classical level in (SUSY) AdS vacua \cite{Derendinger:2004jn, DeWolfe:2005uu}. One drawback of these constructions is that the backreaction of the intersecting O6 planes, required for consistency, is not understood away from the smeared limit \cite{Acharya:2006ne}. But since these vacua are already at the classical level fully stabilised it cannot be that quantum corrections can provide a lift, neither does any other controllable uplift source exists that could turn these AdS vacua into dS vacua \cite{Kallosh:2006fm}.

In case of  the heterotic string theories, the results are again scarce. For instance in \cite{Parameswaran:2010ec} some explicit heterotic orbifold models were tried with potentials generated by various quantum effects. This was enough to stabilise all moduli but no meta-stable dS vacua were found. More succesfull claims can be found in \cite{Cicoli:2013rwa} which studies smooth Calabi-Yau compactifications of the heterotic string with all possible ingredients (incl fractional fluxes \cite{Gukov:2003cy}).

\section{They all have problems?}\label{sec:problems}
The zoo of dS constructions that we briefly outlined in the previous section, can be further subdivided into vacua for which SUSY is broken at (above) the KK scale or far below. This is relevant when discussing how trustworthy a construction is. When SUSY is broken well below the KK scale, one could justify a lower-dimensional effective field description that is a supergravity theory where the dS vacuum breaks supersymmetry spontaneously. This constrains the effective action much stronger compared to models that break SUSY at or above the KK scale. We will not discuss this in any detail further on, so let us mention here that the classical vacua typically break SUSY at the KK scale, whereas the ``quantum IIB vacua", where SUSY is broken by anti-branes for instance should have SUSY broken below the KK scale.

\subsection{Classical}
To date there is not a single 4d classical de Sitter solution known that is free of tachyons. The most extensive scan of models \cite{Danielsson:2011au} (order 1000) always gave at least one tachyon among 7 real scalars. There is a simple argument to understand why this is the case \cite{Danielsson:2012et}. Of all ingredients contributing to the scalar potential, the orientifold tension is the manifest negative term. The size of that term depends on the size of the cycle(s) wrapped by the orientifolds. Hence, in a given critical point of the potential, it is very plausible that a slight increase of the cycle volume lowers the energy, signaling a tachyonic direction. However, for this argument to be fully waterproof one must be certain that the dependence of the other terms on the cycle-volume is less strong than the orientifold tension term. While this is not true parametrically,  the argument seems to work in the sense that indeed all models  have a significant fraction of the tachyonic field direction along the orientifold volume \cite{Danielsson:2012et}.

Further investigations of this matter \cite{Junghans:2016uvg, Junghans:2016abx} have tightened the constraints on stability further, but a full proof that classical dS are always be tachyonic is not (yet) found. Unless one suspects there to be some conspiracy against the existence of dS solutions, one could try to argue that the tachyons are a consequence of statistics of random multi-variable functions. It is a simple fact that the number of local minima is exponentially suppressed by the number of variables (scalar fields) \cite{Denef:2004cf, Marsh:2011aa, Chen:2011ac, Sumitomo:2013vla}.

A further technical problem with the classical solutions is the smearing of the O6 sources. The smeared approximation allows one to verify that the critical point of the 4D potential solves the 10D equations of motion \cite{Acharya:2006ne, Danielsson:2009ff, Danielsson:2010bc}. The problem is not that the solutions can only exist when the sources are smeared, rather it is not understood how to find solutions with fully localised sources (when they are intersecting).  This problem is related but not identical to the understanding of how to include the warping into the effective field theory. Smearing orientifold sources might seem dramatic, but it does not need to be. One simply course grains over length scales of the extra dimensions: the backreaction of the sources is taken into account in an averaged sense where the delta-function sources are replaced with smooth form distributions. 

At least for the known orientifold flux compactifications of the no-scale Minkowski type, the smeared limit and its comparison to the localised solution is well understood \cite{Blaback:2010sj}. Nonetheless this procedure has been criticized, especially when one smears over negatively curved extra dimensions \cite{Douglas:2010rt}. It is the authors opinion that this criticism is in part incorrect since many examples with localised solutions, that have a well-defined smeared limit exists, also with negatively curved dimensions \cite{Blaback:2010sj}. We do not imply that smearing is therefore harmless. One obvious problem that arises with smearing is that it can wipe out instabilities that otherwise are fatal. Imagine a non-BPS brane setup, like a brane-anti-brane pair. Clearly one cannot smear this since it create effectively a brane without charge that is seemingly stable, whereas it is not. Any instability that would manifest itself by some motion inside the extra dimension would be inconsistently course-grained over.
However, whenever the smeared source is still calibrated one could expect the smearing to be consistent and this is the case with the intersecting O6 planes that have been used in the IIA models. 

Yet another worry, not unrelated to smearing sources, is the use of a 4D effective supergravity description because supersymmetry is broken by the combination fluxes and sources such that the supersymmetry-breaking scale can easily be at or even above the KK scale. This is not a problem per se for establishing the existence of a de Sitter solution since most models used consistent truncations (at least in the presence of smeared sources), but it is problematic for understanding the stability of the light degrees of freedom.

Other more stringy worries, such as the understanding of O6 planes in the presence of Romans mass, have been raised \cite{McOrist:2012yc} as well.

In summary, the most-pressing problem of classical dS vacua is the presence of tachyons and the use of four-dimensional supergravity as an effective description. 
 
\subsection{Non-geometric}

Non-geometric fluxes are usually considered as dangerous ingredients whose consistent use in 4D supergravity is not fully understood. One could hope that vacua that fail to be locally geometric, including existing cases with metastable dS, could still be captured by certain 4D supergravities. Those supergravities should describe both string momentum and string winding modes. Therefore some of the extra dimensions must be close to the self-dual radius in order for winding modes and momentum modes to be relevant at low energies at the same time.  This immediately raises a puzzle since it violates one of the assumptions we made before (large enough dimensions). Therefore it is natural to expect that full-blown string theory becomes necessary to understand such backgrounds and, secondly, there can be an infinite KK tower that does not decouple. This is a severe point that we have not seen fully addressed in the literature, although  some progress towards understanding supergravities from non-geometric fluxes using worldsheet techniques has been reported in \cite{Condeescu:2013yma, Blumenhagen:2014gva}.

But one should be careful in what is called geometric and what is not. Among the genuinely non-geometric vacua - meaning that they are not on some T-duality orbit of a geometric vacuum - one needs to distinguish between those that are locally geometric and those that are not. In the language of double field theory, where separate coordinates are introduced for momentum and winding modes, locally geometric theories admit a description where the fields depend on only half of the coordinates \cite{Andriot:2012wx}. In this way one regains a geometric description, even though there are no global dualities that map the theory onto one with only geometric fluxes. Sometimes such theories can be understood as arising from compactifications on non-trivial topologies, with M-theory compactified on an orbifold\footnote{In this discussion we ignore the twisted moduli sector that originates from the orbifolding and we refer to \cite{Dibitetto:2012ia} for the details of the orbifolding and the truncation to minimal supergravity.} of $S^7$ as the most well known example. 

We know consider the STU-truncations of the supergravity actions for M-theory compactified on a (orbifolded) 7-dimensional twisted torus. The fluxes that are geometric contribute to the effective superpotential as follows: 
\be
W_{T^7}=a_0-b_0 S + 3 c_0 T -3 a_1 U + d_0 ST - c'_3 T^2 +3 c_1 TU + 3 b_1 SU +3 a_2 U^2 \,. 
\ee
Here $a_0$ corresponds to the $F_7$ flux, $b_0, c_0$ and $a_1$ to the $F_4$ flux, while the remaining terms come from the curvature of the compact dimensions (sometimes called metric flux). Non-geometric fluxes would correspond to terms that are cubic terms, and higher, in the scalars $S, T$ and $U$. 

As observed in \cite{Danielsson:2014ria}, the $T^2$- and $ST$-terms correspond to non-geometric fluxes in type IIA, but become geometric when uplifted to M-theory. Reversely, the Romans mass, $U^3$, which is geometric in type IIA, fails to be geometric in 11 dimensions. As reviewed in \cite{Danielsson:2015tsa}, the superpotential for M-theory on an (orbifold of) $S^7$ is given by 
\be
W_{S^7}=a_0-b'_3 ST^3 -3c'_1 T^2 U^2-3 d'_2 STU^2 ,
\ee
which lies on a different duality orbit from $W_{T^7}$ of the twisted tori. The fact that these fluxes formally appear to be non-geometric, is just a consequence of forcing a theory with non-trivial topology, i.e. $S^7$, onto a twisted tori. In \cite{Danielsson:2015tsa} it is shown that type IIA with Q-flux generates compactifications on $S^4 \times T^3$, while in \cite{Danielsson:2015rca} it is argued that type IIB with P- and Q-flux gives rise to $S^3 \times S^3$. 

Unfortunately, none of the examples with metastable dS-vacua found seem to be locally geometric, and consequently they lack a geometric reinterpretation of this kind. On the other hand, there is no proof that such examples can not exist.

It is interesting to note that many dS-vacua obtained using non-perturbative additions, such as $P (S,U)e^{\alpha i T}$, with $P (S,U)$ a polynomial, can be captured using non-geometric fluxes \cite{Blaback:2015zra}. The non-perturbative features will be different, but local properties such as the existence of a critical point, as well as moduli masses, can be fitted, since these depend only on the first couple of orders in an expansion. In this way, non-perturbative vacua and vacua based on non-geometric fluxes are, for practical purposes, 
phenomenologically equivalent. Similarly, our knowledge of what actually makes sense or not is, as we have argued, at about the same level. This leads us to conclude that we are dealing with no more than two different ways of parametrizing our ignorance.

\subsection{Quantum}
The difficulty with de Sitter constructions that rely on (quantum) corrections to the tree-level 10-dimensional supergravity description is twofold. On the one hand, it is difficult to explicitly compute derivative corrections, string loop corrections and non-perturbative corrections. On the other hand, even if these can be computed, it is difficult to understand when the corrections are relevant and how the tower of corrections can be truncated self-consistently. In other words, we face the well-known physics slogan \cite{Denef:2008wq}: when corrections are important they cannot be computed and when they can be computed they tend to be not so important. The art of dS model building (or moduli-stabilisation in general) consists in finding the delicate balance of incorporating certain leading quantum corrections and ignoring the rest. 

Despite these difficulties, this method of moduli stabilisation has been given most attention. The reasons for that are most likely due to the common belief that evading the nogo-theorems \cite{Maldacena:2000mw} is most easily done using quantum corrections, although negative tension objects (orientifolds) are sufficient. Another reason is probably that, when relying on 3-form flux compactifications of IIB, one can keep on using (conformal) Calabi-Yau geometries, for which many tools are available and a good understanding of the moduli has been achieved.

We now highlight 3 problems with the IIB constructions that have been pointed out in the recent literature: 1) Issues with anti-brane backreaction on the internal geometry \cite{Bena:2009xk, Bena:2012ek,Bena:2014bxa, Bena:2015kia, Bena:2016fqp, Blaback:2012nf, Gautason:2013zw, Danielsson:2014yga, Michel:2014lva, Danielsson:2016cit, Cohen-Maldonado:2015ssa, Cohen-Maldonado:2016cjh}, 2) issues with anti-brane backreaction on the 4D moduli, which was found recently by Moritz, Retolaza, and Westphal \cite{Moritz:2017xto} and 3) issues with very basic assumptions of the moduli-stabilisation scenario as discussed by Sethi \cite{Sethi:2017phn}.

\subsubsection{Anti-brane backreaction inside extra dimensions}
Since 2009, with the work of \cite{Bena:2009xk} (and \cite{McGuirk:2009xx}), there has been an extended debate about the consistency of the probe approximation used for describing the uplift-effect of $\overline{D3}$-branes. We briefly summarize this debate here, postponing a genuine review (with proper referencing) to a future work \cite{D3review}. 

The use of the probe approximation  means that the uplift energy and perturbative stability is computed from the worldvolume action of the $\overline{D3}$-branes evaluated in a background that is unaffected by the presence of these branes. Going beyond that approximation would imply a study of the back-reaction on the local throat geometry. This is a daunting task given that explicit (compact) CY metrics are not even known, let alone finding solutions once SUSY is broken by $\overline{D3}$-branes. A possible approach to study the effects of backreaction could be to set-up an effective field theory description allowing for a perturbative series expansion, with as leading order term the probe approximation. Some initial steps were described in \cite{Michel:2014lva}, whereas a systematic procedure should probably be carried out in the framework of the blackfold approach \cite{Emparan:2011hg, Armas:2016mes}. 

Instead, most studies sofar have made certain simplifications in order to solve the supergravity equations of motion.  The study of the resulting solutions should give qualitative and quantitive insights on the effects of backreacting $\overline{D3}$-branes. These simplifications were:
\begin{itemize}
\item To study $\overline{D3}$-branes in non-compact geometries with explicit metrics. The canonical choice is then the Klebanov-Strassler solution, which serves as a good approximation to throats in compact CY spaces \cite{Giddings:2001yu}. Working non-compactly also implies one does not need to worry about moduli-stabilisation and one can simply study $\overline{D3}$-branes inside the throat without having to worry about the runaway of the volume modulus in absence of the quantum corrections (since we use 10D supergravity). This non-compact set-up is of separate interest to holography, where $\overline{D3}$-branes inside the KS throat were argued to be dual to dynamical supersymmetry breaking in the KS gauge theory \cite{Kachru:2002gs, DeWolfe:2008zy, Bertolini:2015hua, Krishnan:2018udc}, on the condition that the $\overline{D3}$-branes are at least meta-stable. 

\item One can then further smear the $\overline{D3}$-branes over a compact subspace of the throat (the A-cycle). The effect of this is that the  resulting supergravity equations of motion become coupled ODE's, which is a great simplification.

\item One can linearize the resulting ODE's by setting up a perturbative expansion in terms of the SUSY-breaking ``order-parameter'' $p$, which counts the $\overline{D3}$ charge.
\end{itemize}  
The resulting system of equations was impressively solved in \cite{Bena:2009xk} and follow-up papers. The source of the debate/worry about KKLT stability came from the observation in \cite{Bena:2009xk} that the solution, within the approximations made, contains an unphysical singularity near the $\overline{D3}$ at the tip of the throat geometry. This singularity is not the standard Coulomb-like singularity in the $F_5$ field or metric, but instead it is a singularity in the 3-form fluxes $F_3, H_3$ that deviate in a singular way from the original ISD configuration. 

Such a singularity was argued to induce a perturbative instability that makes the $\overline{D3}$ branes annihilate against the surrounding 3-form fluxes \cite{Blaback:2012nf, Danielsson:2014yga}. It is therefore crucial to understand whether the effect could be a consequence of the approximations made, as suggested in \cite{Dymarsky:2011pm, Baumann:2014nda}. This was shown not to be the case in \cite{Gautason:2013zw} (and \cite{Blaback:2014tfa}). Hence a natural question is then whether the singularity is somehow an artefact of using supergravity instead of full string theory. This was indeed argued for a single $\overline{D3}$ brane \cite{Michel:2014lva, Polchinski:2015bea}. However, things do not need to be too complicated and supergravity can be smart enough: reference \cite{Cohen-Maldonado:2015ssa} (and \cite{Cohen-Maldonado:2016cjh}) demonstrated that the singularity can be removed once the $\overline{D3}$ brane is replaced by a spherical NS5/D5 with non-zero worldvolume flux, exactly as predicted to happen in the probe approximation \cite{Kachru:2002gs}. This effect is known as brane polarisation or the Myers effect \cite{Myers:1999ps}. Let us briefly outline the computation behind this result as it can be understood in a rather straightforward way. 

Similar to Smarr relations for black holes, there exist Smarr relations for branes in flux backgrounds. For the case at hand, $\overline{D3}$-branes in KS, one can derive the following relation (at zero temperature) \cite{Cohen-Maldonado:2015ssa}:
\begin{equation}\label{smar}
M =  \Phi_{\overline{D3}}Q_{\overline{D3}} + \Phi_{NS5}Q_{NS5}\,, 
\end{equation}
where $Q_{\overline{D3}}$ is the $\overline{D3}$-charge and $Q_{NS5}$ the dipole NS5 charge in the system. The potentials $\Phi_{\overline{D3}}$/$\Phi_{NS5}$ are the components of the $C_4/B_6$ potential along the $\overline{D3}/$NS5 horizons. Finally $M$ is the generalised ADM mass of the space-time, which measures the energy with respect to the SUSY KS vacuum and was first computed in \cite{Dymarsky:2011pm} using the UV asymptotics of the ``Saclay solution" \cite{Bena:2009xk}. The key property of this equation is that it relates information measured in the UV, namely the ADM mass, to information in the deep IR, namely the potentials at the brane horizons. The way this equation helps in understanding the existence of singularities comes from one extra identity that can be derived \cite{Cohen-Maldonado:2015ssa}:
\be\label{flux}
e^{-\phi}H_3^2|_{IR} \propto  \Phi_{\overline{D3}} e^{-10A}|_{IR}\,, 
\ee 
where this equation needs to be used in the IR near the anti-branes, where $e^{A}\rightarrow 0$. Hence this clearly shows that the three-form fluxes will diverge unless $\Phi_{\overline{D3}}=0$. This is sufficient to understand the presence of divergences because all supergravity solutions that were considered in the literature did not take into account the polarisation into 5-branes. In other words, all papers used $Q_{NS5}=0$. But clearly, the Smarr relation (\ref{smar}) then informs us that the only way to have positive energy from SUSY-breaking is by having $\Phi_{\overline{D3}}\neq 0$, which causes the singular flux via (\ref{flux}). Once the supergravity Ansatze allow for dipole charge one can consistently take $\Phi_{\overline{D3}}=0$ and divergences should be avoidable. Note that this is entirely expected from the probe result \cite{Kachru:2002gs}, where it was shown that a pure $\overline{D3}$ is perturbatively unstable against decay into NS5 shells. Hence a consistent (ie regular) sugra solution should have dipole charge.

This however does not settle the story. Perhaps the singularity was a red herring\footnote{The singularity is not a red herring in other circumstances. For example, $\overline{D6}$ branes have a similar singularity that does not get resolved by brane polarisation and which can be shown to lead to the predicted instability against brane-flux annihilation \cite{Danielsson:2016cit}.}, but it led researchers to start digging in the right spot. For instance, the 3-form fluxes do not become singular near the branes anymore, but they are still expected to ``clump'', because they carry opposite charges with respect to the branes. Hence if this clumping is too high instabilities can still occur. Other potential sources of instabilities are reported in \cite{Bena:2016fqp, Bena:2015kia, Bena:2014jaa, Danielsson:2015eqa}, although the instabilities of  \cite{Bena:2016fqp, Bena:2015kia, Bena:2014jaa} remain difficult to interpret in our opinion since they cannot be seen as brane-flux decay. Therefore it is not clear whether they are truly harmful and maybe they simply correspond to the brane-shells having to reposition inside the deep throat \cite{Michel:2014lva}.

 Finally, there exists a worrying interpretation of the Smarr relation (\ref{smar}), explained in \cite{Danielsson:2016cit}, which states that the total energy equals the \emph{on-shell} probe brane actions \cite{Gautason:2013zw, Ferrari:2016vcl}. In our case, the right hand side of (\ref{smar}) equals exactly the on-shell Wess-Zumino term since the DBI term is absent. This is worrying because NS5-stability is obtained by letting the DBI forces compete with the WZ forces and exactly in the case of the instabilities conjectured in \cite{Blaback:2012nf, Danielsson:2014yga} would one find that the DBI forces redshift away compared with the WZ forces.

Different evidence pointing to the fact that $\overline{D3}$-backreaction can be unexpectedly large can be found  from the effect on the 4D moduli as described in the next section. Typically the concept of backreaction is best seperated into 4D backreaction, which means the effect of the brane on the 4D scalars (how much they shift after the uplift) and 6D backreaction, what we just described. However, once the effects of 6D backreaction are properly integrated over, it should translate into 4D backreaction.

\subsubsection{Anti-brane backreaction on the moduli}

Let us set aside the worries about the stability of the $\overline{D3}$ branes and their local backreaction in the KS throat. 
So we step away from non-compact throats where gravity decouples. Once $M_{pl}$ is finite we have to consider the possibility that there is an interplay between the gaugino condensates that stabilises the volume and the SUSY-breaking. If such an interplay would be non-negligible then it could be that the 4D scalar potential is not simply the addition of the anti-brane tension (times two) to the original potential. This was the 4D interpretation of a 10D computation carried out recently in \cite{Moritz:2017xto}. This 10D computation revealed that a single gaugino condensate together with $\overline{D3}$-branes \emph{cannot lead to dS vacua}, thereby contradicting the original KKLT paper \cite{Kachru:2003aw}. 

This 10D computation can roughly be summarized as follows. Using the assumption, well-motivated in \cite{Koerber:2007xk, Dymarsky:2010mf, Baumann:2010sx}, that the gaugino condensate can be described at the 10D level by including the fermion bilinears in the D7 probe action \cite{Martucci:2005rb}:
\begin{equation}
    S_{D7} \supset\int_{\mathcal{M}_{10}} \delta^{(0)}_D e^{\phi/2}e^{-4A} \frac{\bar{\lambda}\bar{\lambda}}{16 \pi^2} G_3 \wedge \star_{10} \Omega + c.c.\,,
\end{equation}
then a computation of the kind done in the original GKP paper \cite{Giddings:2001yu} results in:
\begin{equation}  \label{master}
    \nabla^2 \Phi^-  = R_{4} + e^{-6A}\left|\d \Phi^- \right|^2
    + \frac{e^{2A}}{\text{Im}(\tau)}\left|G_3^-  \right|^2  + \Delta_{\text{gaugino}}  \,.
\end{equation}
We used the following short-hand notations:
\begin{equation}
    \Phi^{\pm} = e^{4A} \pm \alpha, \quad G_3^{\pm} = \frac{1}{2} (\star_6 \pm i)G_3\,,
\end{equation}
and $\Delta_{\text{gaugino}}$ is the contribution coming from the fermion bilinears.  The term on the LHS integrates to zero on a compact manifold. This means that the RHS also must vanish. In the absence of a gaugino condensate we find the Minkowski vacuum for which $R_4=0$, $\Phi^{-}=0$ and $G_3^{-}=0$. However the gaugino condensate is a source of  $G_3^{-}$
 and one can show that the combination of the last three-terms in equation (\ref{master}) is positive such that $R_4$ is negative. This reproduces (from a 10D point of view) the KKLT AdS vacuum. 
 
 The essential observation of \cite{Moritz:2017xto} is that the addition of $\overline{D3}$-branes adds a manifestly positive term $2N_{\overline{D3}}\mu_3\delta$ to the right-hand side of equation (\ref{master}). Then one again arrives at the conclusion that $R_4<0$ since the positive contribution of the first three-terms did not change too much (and certainly did not change sign).

The 4D interpretation of this remarkable result, presented in \cite{Moritz:2017xto}\footnote{See also \cite{Progress} for a slightly different version.},  is that indeed the $\overline{D3}$ energy does not simply ``add'' to the energy of the SUSY background once we are in a compact set-up. The reason is the backreaction of the $\overline{D3}$-branes on the moduli-positions. Given the importance of this result it would be relevant to add further evidence to the 10D computations done in \cite{Moritz:2017xto} and to its 4D interpretation.

Interestingly a loophole was suggested in \cite{Moritz:2017xto}: if one would use several gaugino condensates, say from seperated stacks of D7 branes, then there exists a fine-tuning that could evade the above reasoning. This fine-tuning is nothing but the well-known \emph{racetrack fine-tuning} first discussed in \cite{deCarlos:1992kox, Kaplunovsky:1997cy} and more recently in \cite{Kallosh:2004yh} in the context of KKLT flux compactifications. 

Racetrack superpotentials are of the form:
\begin{equation}\label{racetrack}
W = W_0 + A\exp( ia T) + B \exp(ib T)\,,
\end{equation}
and allow a fine-tuning of the coefficients $A,B,a, b$ such that the SUSY AdS vacuum prior to uplift is arbitrary close to Minkowski, while preserving finite masses for the $T$ modulus.	After this fine-tuning, sufficiently small SUSY-breaking effects necessarily lead to dS vacua. 

Although consistency of the racetrack models has been questioned from a pure field theory point of view \cite{Dine:1999dx}, they have been very popular in the (string-)pheno literature. Given the difficulty to prove or disprove the consistency of the racetrack superpotential and its fine-tuning, it might be more rewarding to investigate whether the racetrack fine-tuning violates some general principle which is hold dear. This indeed seems to be the case since the racetrack fine-tuning can easily be shown to violate the (strong form) of the Weak Gravity Conjecture extended to instantons \cite{Progress}. So either the WGC does not hold or the racetrack fine-tuning is inconsistent. Given the success of the WGC we expect the latter. 

It is quite interesting that the inconsistency of an effective field theory becomes manifest when coupled to gravity (ie when combined with moduli stabilisation). But it is not clear whether the inconsistency already occurs at the level of the field theory, or only when coupled to gravity. In other words, it could be that there does not exists a compact manifold that allows two stacks of D7 branes at finite distances that undergo gaugino condensation and at the same time is stabilised with fluxes. Alternatively it can be that this set-up can be realised but that the effective superpotential is not of the racetrack type, ie, the inconsistency already arises at the gauge theory level, prior to the coupling to gravity. 

Finally, for general worries about the validity of an effective field theory description of anti-brane uplifting we refer to \cite{deAlwis:2016cty}.

\subsubsection{Issues with non-SUSY GKP solutions?}

Motivated by the nogo results for dS vacua in heterotic string theory, Sethi studied the IIB dS constructions in \cite{Sethi:2017phn}, trying to understand why they would not be subject to similar constraints that arise from various dualitities with the heterotic theories. The outcome of this investigation is the claim that the IIB constructions are suffering a possibly severe problem. 

The rough argument proceeds as follows. The IIB constructions use moduli-stabilisation mechanisms that lean on the classical IIB ``GKP'' solutions which break SUSY at the level of 10D SUGRA by having (0,3) pieces in $G$ leading to a non-zero value for the flux superpotential $W_0 \neq 0$. But once SUSY is broken by fluxes one could worry about the tower of derivative corrections to the 10D action\footnote{This worry is justified since many constructions rely on some leading derivative corrections and ignore others, that are expected to contribute as well at the same order that the corrections that were kept contribute.}.

The difference with the SUSY solutions \cite{Dasgupta:1999ss} is that the latter are expected to survive derivative corrections due to ``magical'' cancellations that are not shared by non-SUSY solutions. Hence there should be classical higher derivative forces on the moduli coming from these corrections such that there is no Minkowski solution in the derivative corrected supergravity theory. If that is true, it becomes puzzling how one can add non-perturbative corrections on top of a runaway potential in such a way that moduli are stabilised. In other words, the classical background that one is correcting with quantum corrections is some moduli-rolling, time-dependent cosmological space-time. It is then unclear whether instanton corrections and loop corrections are understood. It could be surprising that a quantum correction is strong enough to stabilise a classical runaway direction. Although it was not stressed in \cite{Sethi:2017phn} all the arguments seem to apply as well to the AdS vacua prior to uplifting. Since the latter can be supersymmetric, it seems an easier task to investigate further the objections of \cite{Sethi:2017phn} in that context. 


\section{What now?}\label{sec:nodS}

We attempted to present a bird's eye view on the dS constructions in string theory and their associated problems. We have focused on particular cases in more detail, such as $\overline{D3}$-brane uplifting in IIB warped throats and classical dS constructions, because these can be studied more from a top-down point of view. In contrast, it is very hard to establish the consistency of dS vacua entirely from an F-term potential that originates from fluxes and (non-) perturbative quantum effects. 

From this analysis we conclude that string theory has not made much progress on the problem of the cosmological constant during the last 15 years.  There is a general agreement that the presence of dark energy should be an important clue to new physics. So far, string theory has not been up to the challenge. Or to be more precise, string theorists have not been up to the challenge. 

The well-motivated introduction of the anthropic principle and the multiverse, was a big relief. The mathematical standards were lowered, and unconstrained model building could set in exploring a wild and free landscape of infinite possibilities. But beyond this suggestive connection between a possible multiverse and the rich mathematical structures of string theory not much solid results have been achieved. We reviewed some fraction of the mounting evidence that most, if not all of this landscape, is a swampland  and we refer to \cite{Banks:2012hx, Brennan:2017rbf, Bena:2017uuz} for similar lines of thought.

We believe it makes more sense to listen to what string theory is trying to tell us, then to try to get out of the theory what one would like to have. In recent years, especially with the program of the Swampland \cite{Vafa:2005ui, ArkaniHamed:2006dz, Ooguri:2006in, Brennan:2017rbf}, there is luckily a growing community that embraces this idea. Perhaps this program really already made its first prediction: no measurable tensor modes in the CMB. 

From what we have seen sofar, we believe that the most sensible attitude is to accept there are no dS vacua at all because string theory conspires against dS vacua. To disprove this claim requires a single dS construction that is explicit enough for it to be scrutinized to death. Hence, for dS model builders, it might prove much more useful to look for such an example. This example should not obey any interesting phenomenological constraint. It can for instance be a 7-dimensional dS universe with compact dimensions of the same size as the dS length, if that is what it takes to make the example simple enough. Clearly, it should be easier to find something like that, than a dS vacuum that has phenomenological features. In any case, if such an example is to be found, it should be of interest to dS/CFT model builders. 

It is not unthinkable that dS space is simply a space that cannot exist quantum mechanically. Just like the Schwarzschild solution is only a static solution at the classical level because of Hawking radiation, similarly dS space might be a self-destroying manifold such that the dS isometries are broken at quantum level. This has been claimed before in several papers that study QFT in curved space \cite{Danielsson:2004xw, Polyakov:2007mm, Polyakov:2009nq, Polyakov:2012uc, Markkanen:2017abw, Bautista:2015wqy, Bautista:2015nxc, Pimentel:2018nkl} and if those papers are correct, then doing a proper string computation should reveal that dS vacua cannot exist.

\subsection{What about dark energy?}

If indeed metastable de Sitter vacua of sufficient life time cannot be found in string theory, there are a number of possible interpretations for actual cosmology.  It could be that string theory is simply wrong, as a theory of our universe, or at least incomplete, as it fails it first experimental test. Another possibility is that we have misinterpreted the observations, and that no long term acceleration is actually taking place. There may also be other possibilities where string theory is made compatible with observations in a more subtle way.

For instance, our constraints on how the dark energy is varying over time is not so strong and a quintessence-like scenario \cite{Caldwell:1997ii} cannot be excluded. Still, quintessence is not only notoriously difficult on its own because of the existence of a light scalar, but it is also not easy to find stringy embeddings \cite{Kaloper:2008qs}. In fact, if there is a conspiracy against de Sitter vacua in string theory, it is not unreasonable that the aversion extends to more general slow roll scenarios. Yet another possibility is that the decay of vacuum energy is not due to scalar field dynamics, but rather intrinsic IR quantum effects \cite{Polyakov:2007mm, Polyakov:2009nq, Polyakov:2012uc} as mentioned before.  Yet another option is that there is no dark energy and we observe the effects of spatial inhomogeneities \cite{Buchert:2007ik,Buchert:2011sx,Buchert:2015iva}.

Alternative constructions, circumventing some of the problems we have discussed, include brane world scenarios. Such higher dimensional implementations can also be given a holographic interpretation, where the presence of gravity, as well as a cosmological constant, are associated with RG-flows \cite{Progress2}. More generally, one could envision models where, contrary to popular assumptions, the quantum effects separating the bare cosmological constant from the one measured at large scales are strong. We find it important that effort is spent on investigating these other exciting, and conceptually challenging, possibilities.

\subsection{What about the AdS landscape?}

Perhaps a more modest goal is to first achieve full control over AdS vacua with stabilised moduli and parametric separation between the KK and Hubble scale, before addressing dS vacua. Since the majority of the claimed dS vacua rests anyhow on small supersymmetry-breaking corrections to this AdS landscape, it is equally relevant for addressing dS vacua. 

It is often thought that an AdS landscape in string theory has been established long time ago, but this is not correct. It is true that Freund-Rubin vacua are easily found and come in infinite discrete families, but we use the word \emph{AdS landcape} in a different way. We let it refer to the set of AdS vacua that have a small cosmological constants when measured in KK units: $L_{\Lambda}/L_{KK}>>1$. As we emphasized earlier, this is a necessary condition for a vacuum to be perceived as lower-dimensional and is the most obvious constraint any phenomenological vacuum should obey. But as stressed before, a vacuum with a separation of scales between the KK scale and the AdS scale is by definition a vacuum with an ``unnaturally'' small cc since the KK scale is a scale of new physics. Hence it is quite a non-trivial achievement, from a field theory viewpoint, to establish the full consistency of such vacua. It should therefore not come as a surprise that this is not easy either in string theory. .

Interestingly, the same nogo-theorems (ie Maldacena-Nunez \cite{Maldacena:2000mw}) that can be used to exclude dS vacua can be extended to exclude AdS vacua with scale separation \cite{Gautason:2015tig}. This means that at least orientifold planes are needed to find an AdS landscape. It is fair to say that, similarly to dS solutions, it is hard to establish rigorously the existence of such vacua. Perhaps the simplest set of vacua are the ``classical AdS" (supersymmetric) vacua in massive IIA with intersecting O6 planes \cite{DeWolfe:2005uu} (see also \cite{Derendinger:2004jn}). It is the only set of vacua\footnote{Aside the ``sister constructions'' in IIB O5/O7 solutions that can be found using T-dual ingredients \cite{Petrini:2013ika, Caviezel:2009tu}. } known to us, where there is a flux number $n$ (4-form flux quantum), that is unconstrained by tadpole conditions, that leads to arbitrary strong scale separation and weak coupling in the limit $n\rightarrow \infty $. If these classical AdS vacua in IIA are truly consistent then this forms a class of solutions for which the density of vacua diverges for arbitrary small cc in KK units (by cranking up the number $n$).

 But as we mentioned before, some doubts about the consistency of these vacua has been raised \cite{McOrist:2012yc, Saracco:2012wc} (for instance because of the smeared orientifolds) and it would be most interesting to investigate the consistency of these vacua further. The lack of an 11-dimensional picture, due to the Romans mass, complicates matters.

Perhaps the most interesting way to check the consistency of an AdS landscape would be to first prove the existence of a CFT landscape with the right properties such that the CFT's are candidate duals to AdS vacua with scale separation \cite{deAlwis:2014wia, Polchinski:2009ch}. This is an open avenue for research, and given the recent progress in understanding the spaces of consistent CFT's one can be hopeful.

A general criticism applicable to many vacua obtained from supergravity, is that they do not include open string degrees of freedom. When such modes are added, the rules of the game change and crucial properties of non-supersymmetric AdS vacua are altered in a fundamental way. The AdS$_7$ and AdS$_4$ vacua discussed in \cite{Danielsson:2016rmq, Danielsson:2017max}, are not only perturbatively, but  even non-perturbatively stable in the closed string sector. (For related results, see \cite{Kiritsis:2016kog}.) However, when open string modes are added, things change as the critical points develop new, tachyonic directions. 

This is all intimately related to the WGC. The conjecture asserts, roughly, that in any consistent theory of quantum gravity coupled to a $U(1)$-field, there exists a particle of mass $m$ and charge $q$ such that $m\leq q M_{Pl}$, where $M_{Pl}$ is the Planck mass. This makes it possible for any extremal black hole to discharge itself and Hawking radiate away, hence there will be no stable remnants. In \cite{Ooguri:2016pdq} the WGC was studied in a more general setting, with branes rather than black holes, and it was proposed that it was not enough to saturate the equality, i.e. extremality, to guarantee stability. Only BPS branes are expected to be stable, implying hat any non-supersymmetric AdS obtained as a near horizon geometry would inherit the same instability.  As discussed in \cite{Danielsson:2016mtx}, there is another interesting way to argue in favor of the instability of any AdS based on near horizon geometries. An observer intending to survive more than the characteristic time scale of AdS, must necessarily pass through the wormholes connecting one universe to another. This happens once for every geodesic oscillation in static AdS coordinates. It is well known that these wormholes are fragile objects likely to rapidly close up unless the matter theory is supersymmetric.

To summarize, in many supergravity truncations without string theory, there are hints that any perturbatively stable AdS vacuum is non-perturbatively stable. On the other hand, adding branes and strings, there is some evidence that non-supersymmetric vacua develop open string instabilities. If the open string sector tend to destabilize non-supersymmetric AdS vacua, possibly in a dramatic way, the situation is hardly expected to be any better for dS-space. 

In any case perhaps the best one can expect of non-SUSY AdS vacua in string theory is that they can be meta-stable but not fully stable  \cite{Ooguri:2016pdq}, because there is no symmetry preventing the decay. But meta-stable vacua should not have holographic duals. If they would, then the CFT should posses a time-scale related to the average life-time of the AdS space, which contradicts the conformal symmetry. This can be seen in a more direct way \cite{Maldacena:1998uz}: the probability to observe decay, ie local bubbles of true vacuum, becomes unity towards the boundary. If all of this is correct then one should worry about dS/CFT \cite{Balasubramanian2001, Strominger:2001pn, Maldacena:2002vr} even more.

\subsection{What about dS/CFT?}

It is commonly accepted that dS vacua in string theory can be at best meta-stable \cite{Goheer:2002vf}, hence one would expect that similar to AdS, there are no holographic descriptions of dS space. In this paper we have been more radical and suggested, like \cite{Brennan:2017rbf}, that even meta-stable dS vacua  are in the swampland. Then a dS/CFT duality in the strict sense would certainly be meaningless in a string theory context.\footnote{Alternatively one can regard dS/CFT in a ``bottom-up' fashion as a tool that exhausts the conformal symmetries of the problem. For instance, in the context of inflation, the dS isometries are always a little bit broken, but one can use dS/CFT techniques to make predictions for CMB observables (see for instance \cite{Afshordi:2016dvb, Hertog:2015nia}). This is not what we have in mind here.}

Nonetheless explicit dS/CFT proposals do exist \cite{Anninos:2011ui, Anninos:2017eib}, but they are all formulated within Vasiliev gravity. Vasiliev gravity is unfortunately very different from Einstein gravity. For instance, it has no tower of massive fields and hence evades the problem of imaginary operator weights of the putative CFT dual. Also very peculiar to Vasiliev gravity is that it can be supersymmetrized in dS space \cite{Sezgin:2012ag}, in such a way that one potentially goes around the usual problems \cite{Pilch:1984aw, Witten:2001kn} associated with supersymmetric dS space. Strong indications for this were recently found in \cite{Hertog:2017ymy}. Hence, much like the Ooguri-Vafa ideas for AdS space, it could be that the success of formulating dS/CFT pairs for Vasiliev gravity is due to the existence of a consistent supersymmetric extension that makes dS space fully stable. 

Is this completely disconnected from string theory? Perhaps not. Vasiliev gravity is thought to be the zero-tension limit of string theory, the opposite to the supergravity limit \cite{Isberg:1993av, Gustafsson:1994kr}. Interestingly, Hull suggested some time ago that there could be exotic string theories with supersymmetric dS vacua, which can be obtained fom time-like T-duality from the ordinary string theories \cite{Hull:1998vg}. Not surprisingly the 10D supergravity description contains field with wrong sign kinetic terms, consistent with standard no-go theorems of supersymmetric dS space in supergravity \cite{Pilch:1984aw, Witten:2001kn}. However Hull suggested that these ghosts are an artefact of the supergravity limit and full string theory, with all its massive higher spin states, should somehow make the theory well defined\footnote{Recent work of Dijkgraaf et al \cite{Dijkgraaf:2016lym} suggests a way to make these string theories well-defined and relates them to gauge theories of supergroups.}. If correct, then the tensionless limit should be manifestly a well-defined Vasiliev theory with SUSY dS vacua without any ghost. Exactly consistent with the recent results of \cite{Hertog:2017ymy}. Hence if this can be seen as evidence that the exotic string theories are sensible, without having to take the tensionless limit, then after all string theory will have dS vacua,  even fully stable ones which preserve supersymmetry. Whether these string theories, if consistent, have anything to do with our universe is unclear. They could equally well be some odd points in theory space, just like Vasiliev gravity. 

\section{Conclusion}\label{sec:conclusion}
We hope we have been able to convince some readers that the nature of dark energy in string theory remains an interesting challenge that has not yet been met in a satisfactory manner.  Perhaps string theory is trying to tell us something non-trivial about dS space and the nature of dark energy?  A paradigm shift might be what is needed to make progress in fundamental cosmology. We therefore think that the most natural assumption, at this point in time, is that string theory conspires against the existence of dS space.

\section*{Acknowledgements}
\noindent
We thank  Giuseppe Dibitetto , Jakob Moritz, Stefano Massai, Fridrik Gautason and Vincent Van Hemelryck for useful  discussions.  TVR likes to thank the organisers of the workshop ``Theoretical Approaches to Cosmic Acceleration, Leiden 2017''. Part of this paper is based on a talk given at that workshop.  This work of TVR is supported in part by the FWO Odysseus grant G.0.E52.14N, and the C16/16/005 grant of the KULeuven. The work of UD is supported by the Swedish Research Council (VR).

\bibliographystyle{utphys}
{\footnotesize
	\bibliography{BiblioThomas}}

\end{document}